# Light Dark Matter Search and Spectroscopy – Brief Review and An Experimental Technique

Masroor H. S. Bukhari

Department of Physics,

Jazan University,

Gizan 45142, Jazan, Saudi Arabia.

Email: mbukhari@jazanu.edu.sa

___________________________________________________________________

**Abstract**

    A brief review of light dark matter particles and their spectroscopy is presented, specifically aimed at understanding the interaction mechanisms of axion and Axion-Like Particles (ALP's) with matter and an interim proposal and some ideas for possible detection of these evasive particles. In order to venture into the highly challenging spectral regions and extremely weak signals involved with these searches (especially for the DFSZ axions), a different scheme is proposed departing from the conventional resonant cavity mass scan and heterodyne detection methods. We aim to look for a fixed mass axionic field and concentrate our search at the corresponding resonant frequency and its higher harmonics using a simple phase-sensitive *dc* detection method, which could possibly be helpful in substantially reducing both the hardware, experiment complexities and long run times. The probable mass range suggested in the model presented here begins with $22.5\pm0.5\mu eV/c^2$ mass value, corresponding to a resonant frequency of $5.4\pm 0.12$GHz, going all the way to its two multiples of 90 and $112.5\pm 0.5\mu eV/c^2$ (corresponding to 24.3 and $27.0\pm 0.12$GHz frequencies, respectively), with high probability of finding an axion/ALP around these mass values, if axions exist and couple to photons. We present a comprehensive measurement strategy and spectroscopy technique based upon this model which revolves around a three-stage amplification and phase-sensitive detection scheme to maximize an axion's resonant coupling to the U(1) fields under a coherent Primakoff effect-like interaction. The feasibility of proposed scheme is demonstrated with some calculations, simulations and preliminary tests. This experimental technique and ideas reported here have significant potential to be developed into an extremely sensitive narrow-range dark matter axion/ALP spectroscopy experiment.

*Keywords:* Axions, ALP's, Dark Matter, Primakoff effect, Phase-sensitive Detection.

**PACS:** 14.80.Va, 14.80.Mz, 14.70.Bh

___________________________________________________________________

Scopus ID: 35213052500, ORCID ID: 0000-0003-3604-3152

# 1. Introduction

Light Cold Dark Matter (CDM) [1] has been suggested as the dominant form of the hypothetical dark matter present in the universe, with axions ($A^0$) being the most plausible candidates constituting it. This kind of matter is suggested to be non-relativistic, with much lower velocities ($v \ll c$) as compared to other forms of dark matter, such as fast neutrinos or other constituents of relativistic Hot Dark Matter (HDM), and has extremely weak coupling to ordinary matter and radiation. Axions were initially posited as an emergent solution of the Peccei-Quinn broken symmetry U(1)$_{PQ}$ [2, 3] in the theory of *Quantum Chromodynamics (QCD)*, in the form of pseudoscalar Nambu-Goldstone bosons, which resulted from the so-called *'misalignment'* of the axion field ($\theta$) at or around the onset of QCD phase transition while lying in a minimum of the (so-called *'Mexican hat like'*) U(1) potential. Axions also enjoy a significant credence within the framework of String theory whereby a large number of dark matter axions are produced from strings and domain walls that slowly decay into an ensemble of background cosmic axions [3].

Evinced later by a number of theoretical studies, axions gradually became the most promising candidates for CDM (other than *Weakly Interacting Massive Particles*, or WIMP's) and one of the cornerstones of the standard model of cosmology, although yet to be observed.

A majority of the standard axionic fields was excluded by a number of experimental searches and astrophysical observations, however a form, known as the *"invisible axions"* [3] has hitherto sustained undiminished significance. The invisible axions, and dark matter fields with similar attributes, often named as *Axion-like Particles (ALP's)*, are an important element of all the contemporary cold dark matter searches. However, owing to very little masses and even lower kinetic energies of such axions, the direct searches are precluded and indirect search techniques are sought as a better alternative, especially the resonant or broadband radio frequency cavity based methods.

Owing to their origins within the theory of QCD, axions share a number of underlying symmetries with neutral pions and hence exhibit many similarities to the latter. They can couple to both fermions and bosons mediated by strong or electromagnetic interactions, respectively, including a two-photon interaction in the case of latter.

The interaction Lagrangian for the coupling of an axion to a pair of photons may be described as;

$$\mathcal{L}_{a\gamma\gamma} = -\frac{g_{a\gamma\gamma}}{4} F_{\mu\nu} \tilde{F}_{\mu\nu} \varphi_a = g_{a\gamma\gamma} \vec{E} \cdot \vec{B} \varphi_a \qquad (1)$$

Where $\varphi_a$ is the axion field, $F_{\mu\nu}$ is the electromagnetic field tensor and $\tilde{F}_{\mu\nu}$ its dual, $\vec{E}$ and $\vec{B}$ are the electric and magnetic fields, respectively, and $g_{a\gamma\gamma}$ is the axion-two-photon coupling, which may be expressed as;

$$g_{a\gamma\gamma} = \frac{\alpha}{2\pi f_a}\left(\frac{E}{N} - \frac{2(4+z)}{3(1+z)}\right) \qquad (2)$$

With $\alpha$ as the fine structure constant of electromagnetism, z the ratio of the up and down quark masses, and $f_a$ being the energy scale at which the Peccei-Quinn symmetry breaks (which is also the axion decay constant), the seemingly unfamiliar factors E and N are the consequent electromagnetic U(1) and color SU(3) anomalous axial currents associated with the axion production. The form of this coupling has deep connections with the hadronic origins of axions, especially in the mixing of wavefunctions of the up, strange and down quarks, with those of the axions, which enables axions to couple to photons. Thus, the photon-axion system becomes a pseudo-resonant state which undergoes oscillations, analogous to the well-understood neutrino oscillations and neutrino flavor mixings.

By introducing a dimensionless coupling of order 1, $g_\gamma$, the axion-two-photon coupling is rewritten as:

$$g_{a\gamma\gamma} = \frac{\alpha}{\pi}\frac{g_\gamma}{f_a} \qquad (3)$$

The value of the ratio E/N and the coupling $g_\gamma$ are model-dependent and are taken as 0 and -0.97 for the Kim-Shifman-Vainshtein-Zakharov (KSVZ) or *"Hadronic axions"* model, whereas 8/3 and 0.39 for the Dine-Fischler-Srednicki-Zhitnitskii (DFSZ) or *"Grand Unified Theory"* model, respectively, which are the two of the most sought-after contemporary invisible axion models [4]. The latter model is more realistic as it is dependent on the Standard Model quarks rather than the *a priori* assumption of some new heavy quarks beyond the SM. Although experimentally more challenging, the axions stipulated under this model are the aim of this proposal. A conservative upper limit for the coupling $g_{a\gamma\gamma}$ has been so far set at around a value of less than or equal to $10^{-10}$ GeV$^{-1}$.

Leaving aside all the well-established experimentally known constants, $f_a$ emerges as the most important factor (and unknown) which determines both the axion mass and the coupling strength of axions to photons. Axions are essentially massless, an effective mass is endowed to them after symmetry breaking by the virtue of instanton effects in the QCD vacuum, analogous to the Higgs mechanism and the creation of other pseudoscalar Nambu-Goldstone bosons. Axion mass ($m_a$) is the second most important and related parameter which originates after symmetry breaking in QCD as a function of the PQ symmetry breaking scale, $f_a$. The origins of axion mass and its scale are also dependent on the temperature scale of the primordial universe and whether the onset of U(1)$_{PQ}$ symmetry breaking preceded the inflation era or followed later. The discussion is beyond the scope of this report and is deferred to existing established treatises on the subject [1, 2, 4].

As per the contemporary cosmological models and observations, the coupling $f_a$ has been assigned (but not limited to) some realistic values. As per the DFSZ model, the value of the constant is estimated at $f_a > 0.8 \times 10^7$ GeV (with a corresponding axion mass upper limit of $m_a <$

0.7 eV). However, for the KSVZ model, its range of values is about three to four orders of magnitude higher, roughly around $f_a \sim 10^{10}$ to $10^{11}$ GeV, which corresponds to an expected axion mass range between $10^{-5}$ to $10^{-4}$ eV. The large values of the axion decay constant reflects on its weak coupling to matter.

Theaxion mass is inversely related to its decay constant/coupling $f_a$ as;

$$m_a = \frac{z^{1/2}}{1+z} \frac{f_\pi m_\pi}{f_a} \qquad (4)$$

Here, z is the ratio of the masses of up and down quarks, $m_\pi$ is the neutral pion mass (~135MeV/c²), and $f_\pi$ is the pion decay constant. As can be seen with the equation, the PQ symmetry breaking scale and axion mass are related to each other and determination of the value of one would immediately fix the value of the other.

The mass can be expressed in a compact form as;

$$m_a = 6.2 \times 10^{-6} \left( \frac{10^{12} GeV}{f_a} \right) \qquad (5)$$

The corresponding axion *Compton frequency*, an experimental parameter of fundamental importance, may be expressed as:

$$\nu_a = \frac{m_a c^2}{h} \qquad (6)$$

Using the equivalence between the Compton frequency and mass, the coupling of axionic fields to electromagnetic fields, ie photons, which are easier to achieve and manipulate in an experimental setting, may be a feasible way of indirectly detecting axions.

First proposed by Skivie [5], it was suggested that the presence of a strong magnetic field can substantially enhance the otherwise small coupling of an axionic field to an ambient electromagnetic field in a RF cavity. The interaction is mediated by means of the axion's two-photon vertex under a meson-like coherent inverse Primakoff process (as illustrated by a Feynman diagram in Figure 1) [6];

$$A + Ze \rightarrow Ze + \gamma \qquad (7)$$

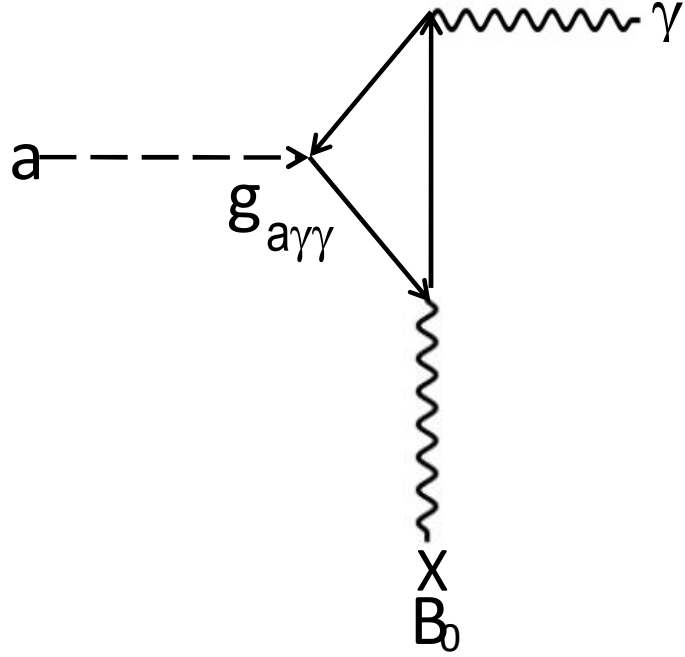

**Fig. 1: The two-vertex axion-photonconversion in a static magnetic field (under a coherent inverse Primakoffprocess)**

In the presence of a high-intensity magnetic field, the coherent inverse Primakoff process-mediated axion-two-photon coupling is associated with a small momentum transfer (***q***) and the interaction remains coherent over a large distance [7], which enables the coupling to take the form of an axion-photon oscillation in analogy to the neutrino flavor mixing oscillations. Thus, it becomes highly probable for such a process to be experimentally detected with sufficiently strong magnetic fields and sensitive measurement techniques.

The conversion probability for the *Skivie axions* in such a setup is expressed as [2];

$$\Pi_{a\gamma\gamma} \sim \frac{1}{4}\left(g_{\alpha\gamma\gamma}Bl\right)^2 \qquad (8)$$

Here, $\vec{B}$ is the intensity of the ambient transverse magnetic field and $l$ is the detector length along which the magnetic field is applied.

The coherence condition yields a range for axion mass sensitivity for a mean axion energy $E_a$ [8] as:

$$m_a \lesssim \sqrt{\frac{2\pi E_a}{l}} \qquad (9)$$

As a consequence, solenoidal cavities with large-scale magnetic fields and sufficiently long coherence lengths offer higher propensity to induce axion-photon mixing.

Building on this theoretical framework, we suggest a model and a corresponding detection strategy for the search of invisible axions, as reported in the forthcoming sections of this paper. The aim of the strategy and our experiment is to maximize the resonant coupling between an incident axion and the electromagnetic modes in the cavity using various improvisations and detect the signal corresponding to it with highest possible precision while increasing the noise overhead.

In the following sections of this paper, we begin, in a pedagogical manner to delineate various important parameters of our model followed by a detailed discussion on the basic building blocks of our proposed experiment and involved instrumentation, with a hope to aid the experimentalists in understanding the issues involved with the axionic searches. Followed by a brief account of the measurement methods, we present the preliminary test results obtained with the proposed scheme with the help of *"synthetic axions"* (artificial axion-like signals) injected by hand into the cavity. Finally, we present a detailed discussion on our results as well as a few important ideas and factors of consideration in realistic axion searches which might prove to be valuable in designing effective axion haloscopes and spectroscopy techniques.

## 2. The Model and Theoretical Estimations

*2. 1 The Model:*

The parameters of our axion model (following both the DFSZ and KSVZ specifications) and detection regimenare enumerated in Table I. Unlike a variable mass scan range technique, a fixed axion mass search is devised here, searching only for some probable and high likelihood values of axion mass.

Based on strong theoretical and cosmological motivations, a minimum axion mass value of 11.25 μeV was contemplated, followed by its multiples as 22.5, 33.75, 45, 56.25, 67.5, 78.75, 90, and 112.5μeV. These mass values correspond to the associated axion Compton frequencies of 2.7, 5.4, 8.1, 10.8, 13.5, 16.2, 18.9, 21.6, 24.3, and 27 GHz, respectively (the initial value $f_0$ and its nine higher harmonics or multiples). However, in view of the exclusion of some of these values by existing experimental searches or cosmological constraints, the most probable values for the axion mass suggested by us are the eight values from 22.5 to 112.5μeV with an error of $\pm$0.5μeV (corresponding to frequencies from 5.4 to 27 GHz, respectively, with an error of $\pm$0.12GHz). Out of these, there are four values which have been chosen as the most probable values of the axion mass, which our search will be concentrated upon, viz. 22.5, 56.25, 90 and 112.5μeV, with confidence that if axion existed, it could be found around one of these values.

| Parameter | Value/Range |
|---|---|
| Mass ($m_a$) | 22.5 $\pm$ 0.5µeV to 112.5 $\pm$ 0.5µeV |
| Corresponding Resonant Frequency (Axion Compton Frequency) ($\nu_{res}/\nu_c$) | 5.4 $\pm$ 0.12GHz to 27.0 $\pm$ 0.12GHz |
| Mean Axion Mass ($\langle m \rangle$) | 101.25+0.5µeV |
| Axion Coupling/Decay Constant ($f_a$) | $10^{11}$ to $10^{10}$ GeV |
| Cosmological Axion Model | KSVZ / DFSZ |
| Axion-γγ Coupling ($g_{a\gamma\gamma}$) | $10^{-15}$ to $10^{-13}$ GeV$^{-1}$ |
| Corresponding Compton Wavelength ($\lambda_c$) | 0.05 to 0.01 m |
| Axion Velocity ($v_a$) | $2.3 \times 10^5$ ms$^{-1}$ |
| DM Density ($\Omega_{dm}$) | 0.3 $\pm$ 0.1 GeV-cm$^3$ |
| Axion Density ($\Omega_a$) | 0.323 $\pm$ 0.1 GeV-cm$^3$ |
| Cavity Radius ($r$) | 2.1cm to 4.2mm |
| Cavity Length ($l$) | 100-240cm |
| Cavity Mode (TMφρz) and Form Factor ($C_{\varphi\rho z}$) | TM$_{010}$ ~0.5 |
| Cavity Quality Factor ($Q_0$) | 0.75 - 1.0 $\times$ 10$^5$ |
| Magnetic Field Intensity (B) | 8.0 - 10.0 T |

**Table I: Specifications of our proposed axion model and detection parameters**

As a result of contemporary cosmological models and astrophysical observations, the overall galactic dark matter density around earth $\Omega_{DM}h^2$ is estimated atvalues such as 0.1143 to 0.12 [9] (where $h$ is the hubble parameter). Based on those models, we choose a unanimously accepted and recent [10] value of $\Omega_{DM} \sim 0.3\pm0.1$GeVcm$^{-3}$ as the total DM density for our model.

The average contribution of axions to the overall DM density has been suggested as [2];

$$\frac{\Omega_a}{\Omega_{dm}} h^2 \approx \left(\frac{f_a}{10^{12} GeV}\right)^{\frac{7}{6}} \qquad (10)$$

Using the inverse relation between $m_a$ and $f_a$, this can be alternatively expressed as;

$$\Omega_a \approx \Omega_{dm} \left(\frac{m_c}{m_a}\right)^{\frac{7}{6}} \qquad (11)$$

For our suggested axion mass value, the corresponding axion density around the earth is, thus:

$$\Omega_a \approx 0.3 \times \left(\frac{m_c}{22.5 \mu eV}\right)^{\frac{7}{6}} \sim 0.323 \pm 0.01 \frac{GeV}{cm^3} \qquad (12)$$

An axion with mass $m_a$ and momentum $\vec{p}$ incident on an earth-bound terrestrial detector, whether in the form of a purported *"axion wind"* or as part of a surrounding galactic halo, would have a Lorentz-boosted total energy (with $\beta_a = v_a/c$):

$$E_a = m_a c^2 + \frac{\vec{p}^2}{2m_a}\beta_a^2 \qquad (13)$$

We assume the velocity and direction of axions coming from the galactic halo to Earth as $v_a \sim 2.3 \times 10^5 \text{ms}^{-1}$ orthogonally.

Most of the energy content of the axion is concentrated in its mass and a small value is carried by its kinetic energy. Since the mass of an axion is quite low and they have very little energies, the energy registered in the detector is negligible, precluding the direct detection. An electromagnetic cavity with its resonant frequency tuned to the corresponding mass becomes a viable method by indirectly measuring the resonance produced by an axion-photon conversion, as proposed here and in many searches before.

An important factor to consider here is a dispersion in the axion kinetic energy owing to the time-dependent velocity dispersion ($\Delta v(t)$) of axions coming from the galactic halo and reaching a terrestrial detector. The axion velocities are thus expected to follow an appropriate velocity distribution profile, an important variable to be taken care of in realistic calculations.

The probability of axion photonic decay in a magnetic field ($\vec{B}$) under an inverse coherent Primakoff effect, as per Skivie's model [5], can be written down in a form as;

$$\Pi_{a \to \gamma\gamma} = \frac{1}{16\beta_a}\frac{\alpha}{\pi^3 f_a^2}(g_\gamma eBcl)^2 \left(\frac{\sin\frac{ql}{2\hbar}}{\frac{ql}{2\hbar}}\right)^2 \qquad (14)$$

Here, $\beta_a$ is the Lorentz-boosted velocity for the axions, $l$ is the length of the detector, and $\vec{q}$ is the axion-photon momentum transfer (which is $m_\gamma^2 - m_a^2/2E_a$ *in vacuo*) with $E_a$ the total energy of the axion. The last term expresses the degree of coherence between the two particle wavefunctions involved with the axion-two-photon conversion and is constant for a fully-coherent conversion. Assuming a coherent conversion and weak coupling of axions, one can appreciate from equation (14) that the value of probability for the axion-two-photon conversion is quite small and the most significant factors contributing to it are the axion mass, detector length and the ambient magnetic field.

*2.2 Power Estimation:*

In terms of the magnitudes of electric ($\vec{E}$) and magnetic fields ($\vec{B}$) involved with photons which arise out of the axion-two-photon conversions, based upon the Lagrangian in equation (1), one finds the value of electric fields as;

$$\vec{E} = g_{a\gamma\gamma}\varphi_a\vec{B} \; ; \vec{B} = \frac{\vec{E}}{g_{a\gamma\gamma}\varphi_a} \qquad (15)$$

In a cavity with the electric field vectors point towards axial direction (along the *z*-axis), the electrical field intensity and the power, the variables of our interest, are expressed as;

$$\vec{E} = (0,0,\vec{E_z}); \vec{B} = (0,0,\vec{B_z}) \qquad (16)$$

$$P = (\vec{E_z} \cdot \vec{B_z}) = \vec{E_z}\vec{B_z}\cos\theta \qquad (17)$$

The power registered within the cavity for a particular mode ($\varphi\rho z$) may be expressed as [5];

$$P_{a\gamma\gamma}(\varphi\rho z) = (g_{\gamma\gamma})^2 V Q_L B^2 C_{\varphi\rho z} \frac{\rho_a}{m_a} \qquad (18)$$

This is the most important expression in any cavity-based axion search, as it gives an estimate of the powers involved with an axion signal. Here V and $Q_L$ are the cavity volume and loaded Quality factor, respectively, and $C_{\varphi\rho z}$ is the cavity axion-photon coupling form factor (normalized to 1) which determines the coupling of an axion to a particular mode of the electromagnetic field within the cavity.

The form factor describes the degree of overlap between the electromagnetic field owing to the cavity dimensions and the electromagnetic fields resulting from axion conversion, and is the integral of the dot product of these two fields [6], described as:

$$C_{\varphi\rho z} = \frac{\left|\int d^3x \vec{B} \cdot \vec{E}_{\varphi\rho z}(\vec{x})\right|^2}{B^2 V \int d^3x \varepsilon(\vec{x}) |\vec{E}_{\varphi\rho z}(\vec{x})|^2} \qquad (19)$$

The magnitude of electric field associated with the TM010 mode in z direction is increased manifold in the wake of resonance, due to this particular geometry. However, the resonant frequency of the cavity is solely dependent on the dimensions of it.

It is important to note that albeit the volume of cavity is important for detection, it scales inversely with increasing resonant frequency and so do the surface resistance of the cavity material and the quantum noise of the amplifiers (in addition to challenges in RF engineering inherent with higher frequencies as we cross the threshold of few tens of GHz).

The expected highest signal power from any possible axion events in a realistic axion haloscope, such as ADMX, is estimated at below or around the $10^{-22}$W (around -180dBm) level which is not a challenging strength to measure with existing technology. However it is limited to frequencies in the MHz range. Once the frequencies increase beyond few GHz ($m_a>10\mu eV$), especially for the axions coinciding with the DFSZ specifications, it falls down by several magnitudes. Besides, the increasingly smaller radii of cavities corresponding to > 10GHz resonant frequencies and consequently smaller cavity volumes, decrease both the power and probabilities of detection, making it virtually impossible to detect anything. A number of solutions have been suggested to find a way across this difficulty, which include inserting metallic inserts in large cavities to increase the resonant frequency while keeping the volume large, bunching multiple cavities with smaller radii (corresponding to higher resonant frequencies) in large volume cavities, or altogether changing the cavity geometry from the cylindrical shape to other geometries such as toroidal or re-entrant geometries etc. (as used in particle accelerators). We attempt to take a middle approach once we cross the resonant frequencies higher than 5.4GHz.

*2.3 Signal-to-Noise Ratio:*

The second most important factor other than power is the Signal to Noise Ratio (SNR) of the measurement scheme.

The SNR of a radiometer, or a microwave signal measurement scheme, can be estimated by an adapted form of the Dicke Radiometer Equation [11];

$$SNR = \frac{P}{k_B T}\left(\frac{t}{\Delta f}\right)^{1/2} \quad (20)$$

Here, $P$ refers to the signal power (as expressed in the equation 18), $T$ the collective physical noise temperature of the system (which is a sum of the ambient, amplifier and post-amplification signal processing stages temperature), $t$ the integration time (i.e. the time over which a measured sample is averaged) and $\Delta f$ the integrated bandwidth (i.e. the measurement bandwidth over which a single measurement is made). This equation is simply a consequence of the Central Limit theorem which specifies how the noise temperature measurement uncertainty scales with the square root of the number of samples. Thus, SNR's can significantly be improved by integration over long periods of time. Typically, the integration time is on the order of minutes and varies from $10^3$ to $10^4$ for our contemplated experiments, and the integrated bandwidth is on the order of few KHz (in our data acquisition it is 20KHz, as set by the system parameters and our DAQ). The expected SNR for axion searches is normally taken as equal to greater than 5, which is a challenge, especially for the DFSZ model axion searches. However, after long integration times, SNRs of up to 10 may be achieved in realistic conditions. This proves to be a great benefit to overcome hardware limitations in axion searches.

Figure 2 illustrates our theoretical estimation of power from both KSVZ and DFSZ axion events, plotted above the noise floor, arising from the cavity detector, for a range of center frequencies of 2.7, 5.4, and 8.1GHz in our model, whereas Figure 3 provides the corresponding raw Signal to Noise Ratio's (SNRs) based upon the radiometer formula (without any signal processing or integration). This gives a rough measure of the difficulties faced with detecting true axionic events and heralds the need for unconventional detection techniques.

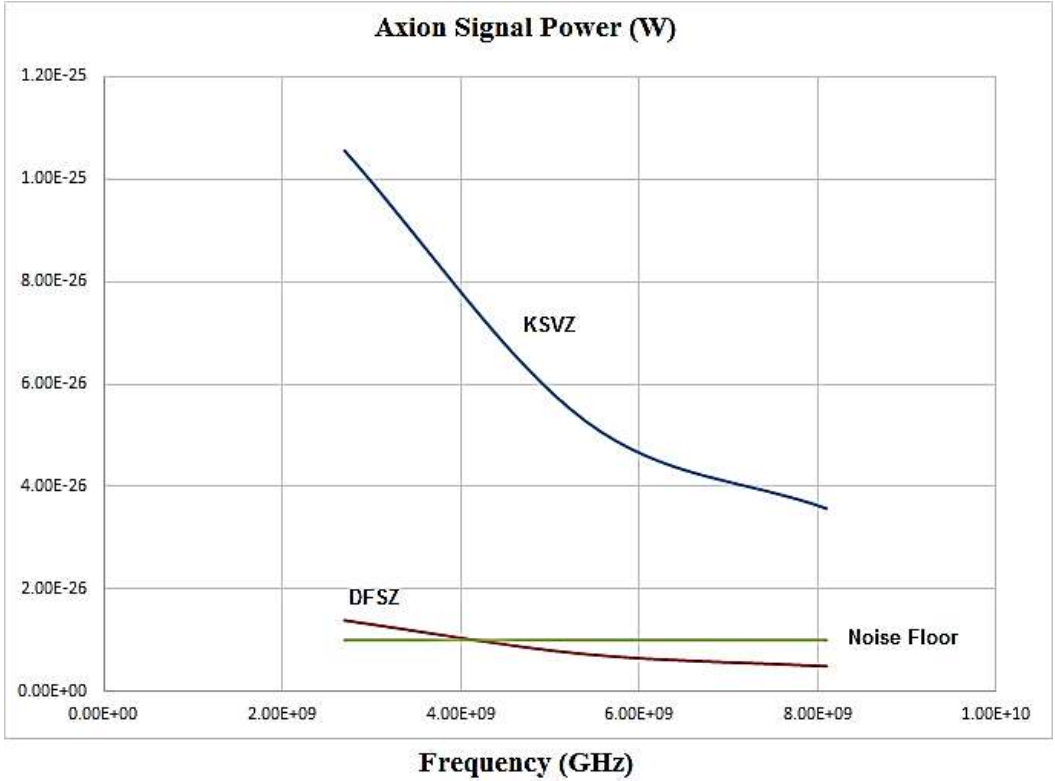

**Figure 2 :** A plot of the estimated signal power of an axion event plotted against the range of frequencies $(f1, f2, f3, ...)$ as per both the axion models (in a discrete manner). We used the values of constants in the equation (14) as follows; $m_a$=11.25 to 33.75µeV, $\rho_a$=0.323GeVcm-3, $g_{\alpha\gamma\gamma}$=0.45 or 3.0GeVcm$^{-3}$ for KSVZ or DFSZ models, respectively, V ~ 2.0 x10$^{-5}$m$^3$, B=8.0T, $C_{\varphi\rho z}$=0.6, $\nu_{res}$=2.7 to 8.1GHz, and $Q_L$=5x10$^5$. The flat line at bottom depicts the noise floor. Note the kinks in the plots do not indicate any inherent non-linearity in power, it is an artifact created because of using discrete frequency values in graph.

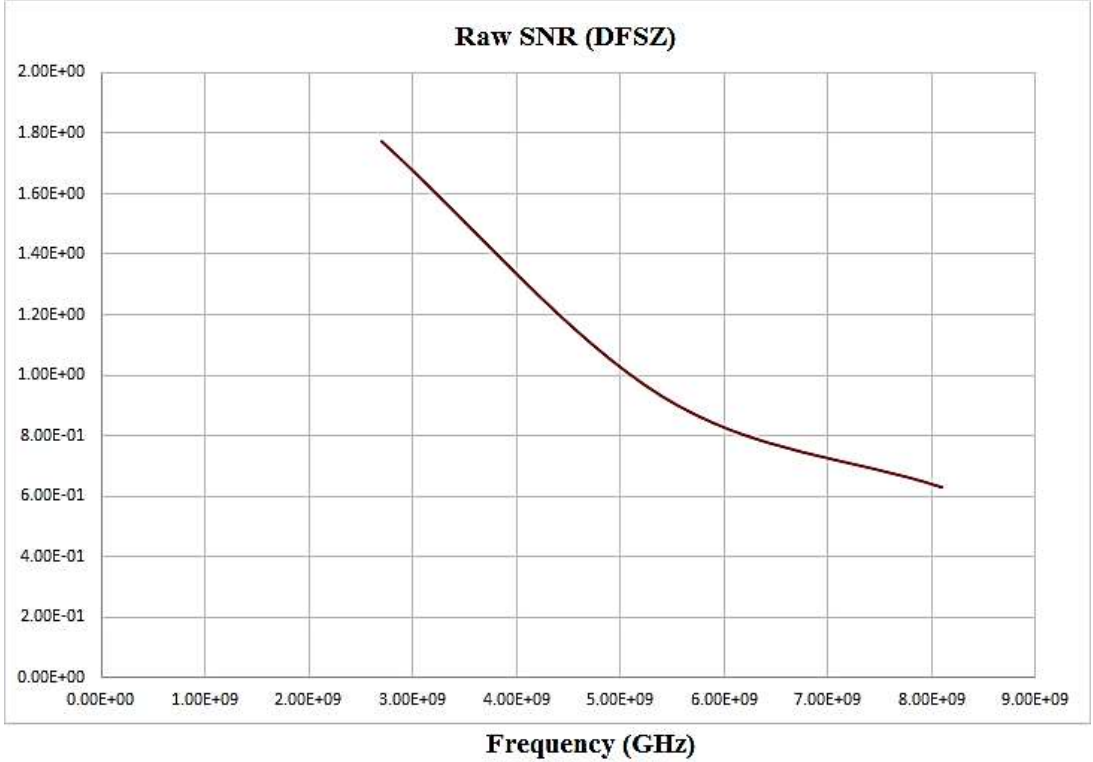

**Figure 3 : Raw Signal to Noise Ratio (SNR) of DFSZ axion events which are the aim of this search proposal. The calculations were done using the Dicke radiometer formula for an ambient physical temperature of the cavity and detector at 20mK and the integration time and bandwidth were taken as $10^3$ s and 20KHz, respectively.**

Based on the equations (6) to (8), the tenet of resonant cavity detection schemes, in general, is to couple an incoming axion to a resonant electromagnetic mode ($\phi\rho z$) in a carefully chosen volume of a high Q-factor cavity permeated by a strong and highly uniform magnetic field, while working at low physical temperatures and taking a large number of samples, over a sufficient but not toolarge a bandwidth, which may facilitate the resonant detection of an axion-photon conversion event beyond the thermal noise. Although the axion signal power depends on a multitude of factors, it has substantial dependence on the strength and uniformity of the ambient magnetic field, the resonant frequency mode and cavity properties (such as volume, finesse and form factor).

The key to successful detection thus seems to lie in reducing the system's noise, concentrating on finding an optimal cavity volume and maximizing its finesse, and obtaining a highly uniform and large magnetic field to maximize axion-photon coupling. Additionally, measurements have to be carried out ideally above the thermal noise floor of the cavity-detector system and amplifier, for substantially long integration times. It is important to note, however, that the loaded Q factor of cavity ($Q_L$) cannot be much larger than the axion signal quality factor ($Q_a$) within the cavity [12] (the two are typically expected to be on the order of approximately

$10^5$ to $10^6$). The secondary factors which aid in the detection strategy are signal processing and analysis techniques, which lie in the domain of software.

In a similar fashion, we attempt to measure a resonant frequency weak signal corresponding to a probable axion-two-photon conversion. An overview of our proposed experiment, which is a based upon the conventional Primakoff effect-based resonant cavity axion detection experiments, is illustrated in Figure 4a. A resonant cavity in a strong solenoidal magnetic field facilitates the conversion of an incident axion into a photon which is detected by an antenna (here we have devised a special tuning fork antenna geometry). The ultra-weak signal is amplified by a three-stage cryogenic amplification cascade which increases the signal power to be read out by a conventional data acquisition and read-out scheme. Figure 4b provides a coarse, false-color, visualization of the z-axis electric field component ($\vec{E}_z$) distribution in the cavity, based on a usual Finite-Element simulation, over which a cartoon of a magnetic field-mediated axion field's conversion into a photon (γ) going outward is depicted.

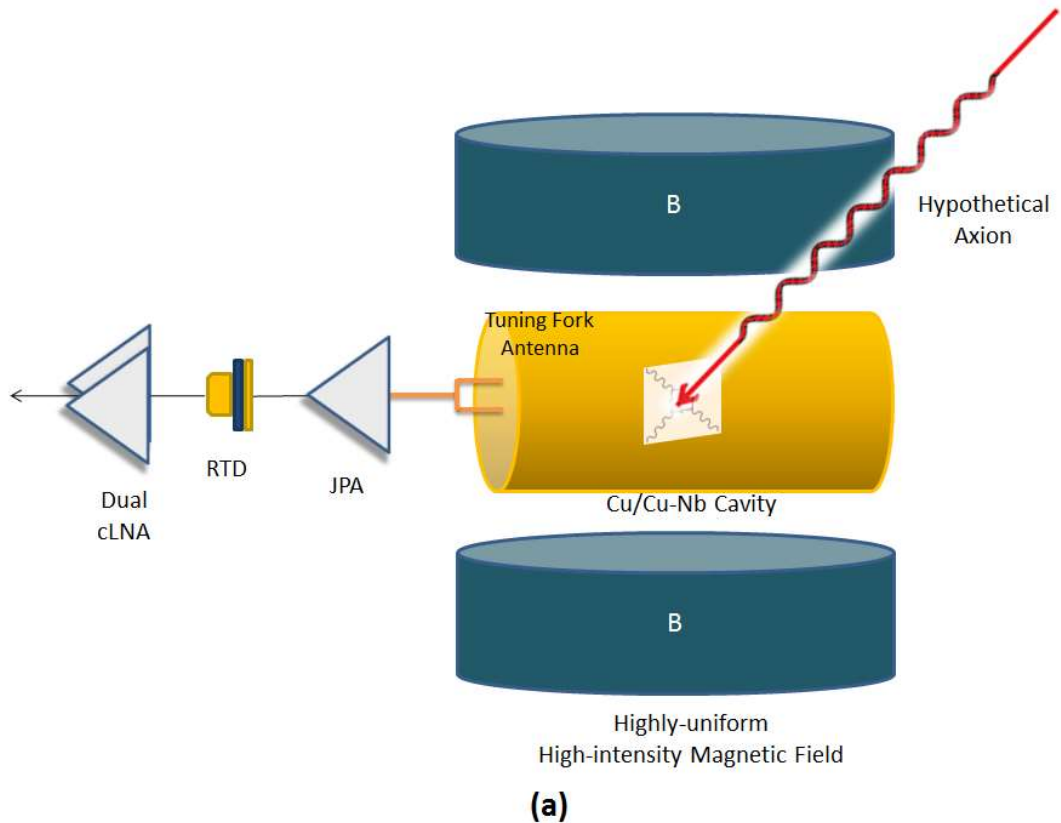

(a)

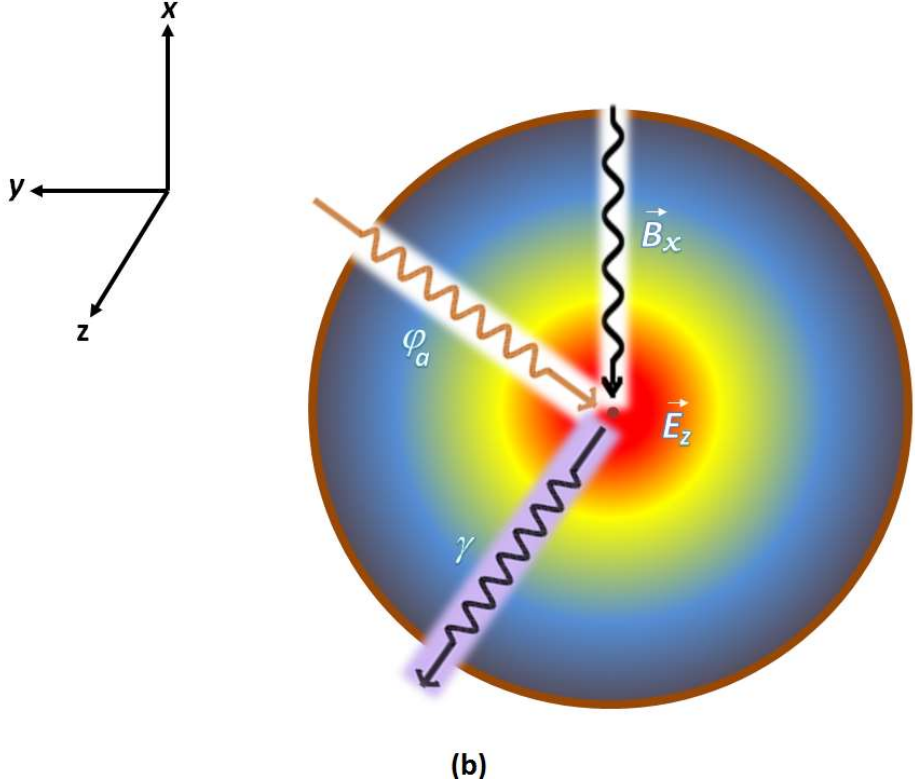

**(b)**

**Fig. 4.** An illustration of the proposed experiment to detect an axion-photon conversion in a resonant cavity under the coherent inverse Primakoff effect in a strong magnetic field. (a) An overview of the experiment whereby a photon generated from an incident axionic field is detected in a resonant microwave cavity operating in a high-intensity magnetic field. A faint microwave signal is generated from an axion-photon conversion in a microwave cavity under the influence of a strong and highly-uniform magnetic field ($\vec{B}$) facilitated by a superconducting magnet at the perimeter of the cavity. A tuning fork antenna critically coupled to the cavity (preferably placed in the center of a waveguide at the perimeter of the cavity endcap to facilitate impedance matching) picks this signal which is amplified by a three-stage cryogenic amplification scheme comprising a Josephson Parametric Amplifier (JPA), a Resonant Tunneling Diode (RTD) and a duo of cryogenic Low-Noise Amplifier's (cLNA) based on High-Electron Mobility Field Effect Transistors (HEMTs/HFETs). An optimal solution would be a specially-fabricated integrated RTD-gated HEMT device, as discussed in text. (b) A cross-sectional view of the cavity illustrating the hypothetical conversion of an axionic field ($\varphi_a$) into an electromagnetic field photon ($\gamma$) under the influence of a transverse magnetic field ($\vec{B}$) in the cavity, resulting into the generation of a photon in the axial direction. The illustration is superimposed on a coarse toy simulation of the $\vec{E}_z$ field distribution in the cavity (the intensity shown in descending order in terms of red, yellow and blue contours, respectively), in the axial direction normal to the magnetic field.

In order to assess the form of axionic signals in cavity and intrinsic noise, including its spectral distribution in the frequency range of our interest, some calculations were performed and simulations were carried out. A 15000-point simulation was written in a computer program and results were accumulated in a data file.

Figure 5 and 6 illustrate the time-domain and frequency-domain results of these simulations, respectively, whereas Figure 7 depicts a histogram of the simulated power's spectral distribution, constructed from the simulation of power as a function of frequency in the 1-12GHz spectral region.

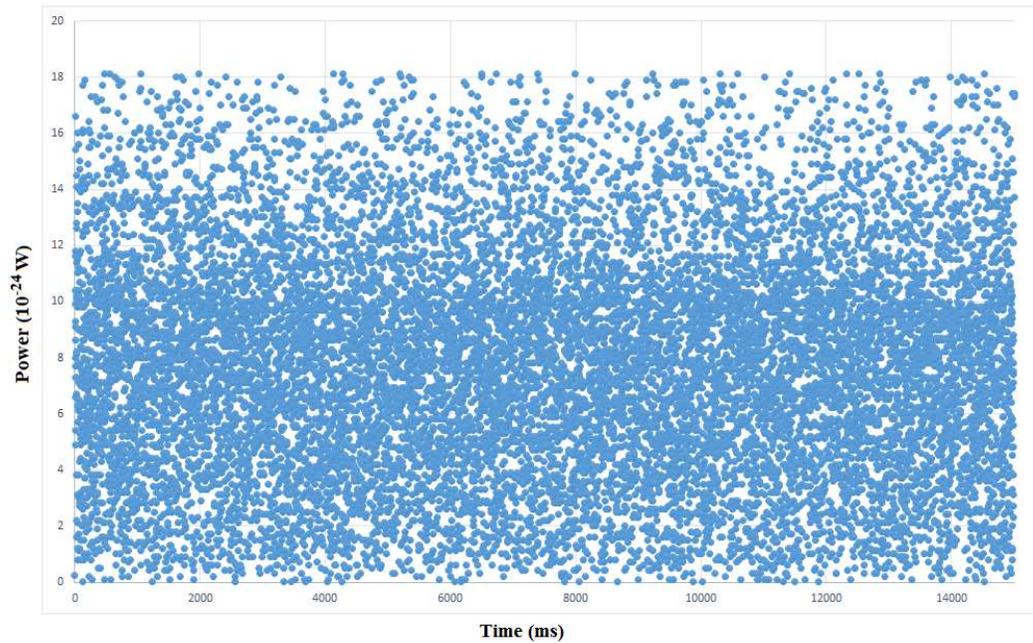

**Figure 5: The result of a 15000-point simulation generated on computer to assess the form of power as generated from axionic events in the cavity in our region of interest, 2-10GHz. The plot gives a time-domain view of the hypothetical signals registered in the cavity volume**

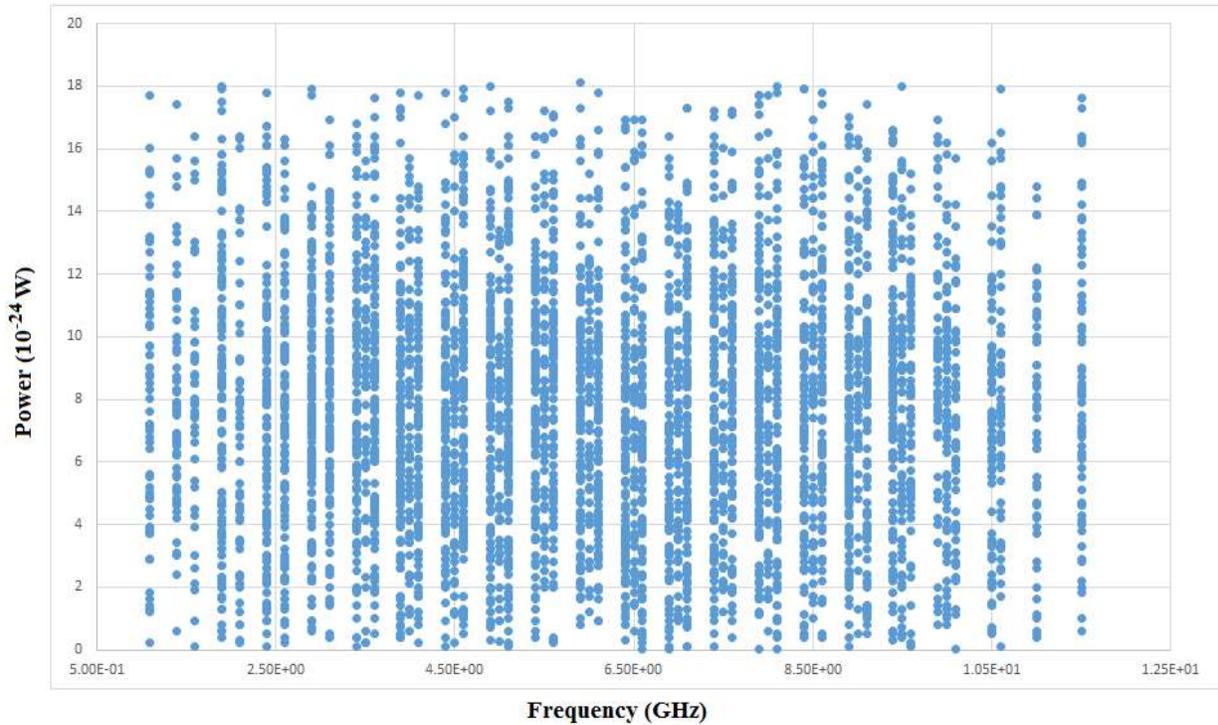

**Figure 6: The spectral distribution of a similar simulation of hypothetical events registered in the cavity volume, however, here as a function of frequency within our region of interest, 1 to 12GHz.**

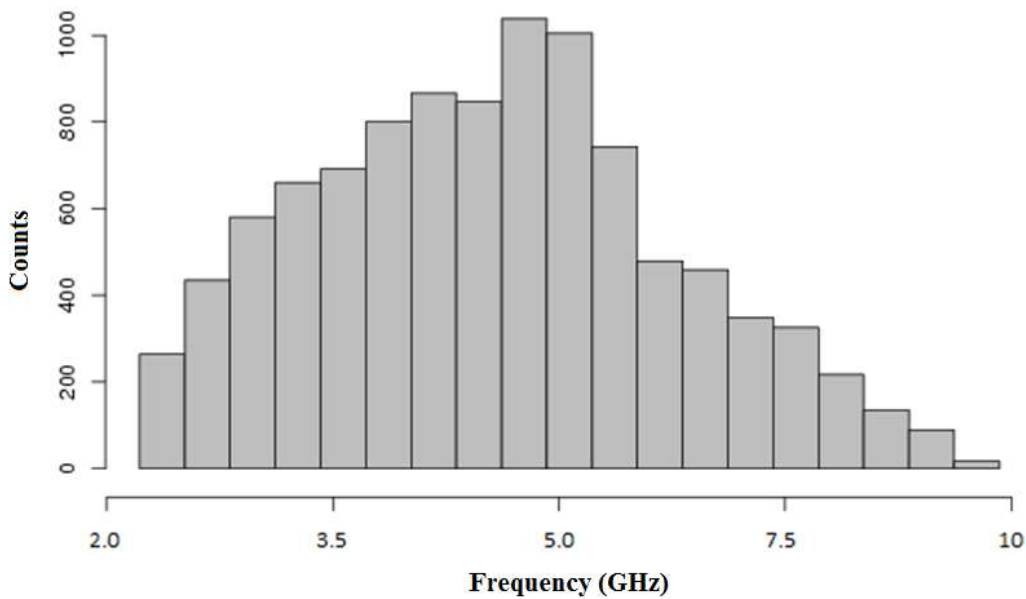

**Figure 7: A histogram depicting spectral distribution of power, constructed from the data generated from the spectral simulation, as illustrated in Figure 6.**

## 3. The Experiment

*3.1 Detection scheme:*

Similar to resonant cavity axion haloscopes, the central part of our experiment is a resonant cavity which provides an environment to facilitate the resonant conversion and detection of an axion/ALP event. The cavities proposed here and being tested are 1.0-2.1cm radius and 50-240cm long cylindrical geometries made with 0.3mm thick oxygen-free 99.99% high-purity copper, with an aim to support the $TM_{010}$ modes (which accord the maximal form factor for a cylindrical geometry). However, we are contemplating cavities of other geometries, especially toroidal cavities, which may possibly have advantages over the existing geometries used in axion searches.

A loaded Q factor of around $1 \times 10^5$ is expected to be achieved with special interior polishing by means of procedures as reported earlier [13]. The cavity is surrounded by a $LN_2$-cooled superconducting solenoid which provides a highly uniform ($\Delta B/B$~1ppb) and homogenous field throughout the interior of cavity. For practical reasons, the magnet sought for the proposed experiments is a 8.0T solenoid.

The detection system architecture comprises three main stages, as has already been illustrated in Figure 4. Figure8 expands on this architecture, illustrating the electronics and instrumentation involved with the experiment in detail. The experiment has three distinct temperature stages, 20mK, 4K and 290K, which are all operated on batteries, without resorting to *ac* power to reduce mains noise and other electromagnetic interferences. The measurement equipment and computers are all at room temperature, operated on 220V*ac* mains supply. For test purposes, the cavities and amplifiers are all kept in a class I solid copper Faraday cage, covered with a special RF protection fiber canopy (Aaronia GmBH). The Lock-In amplifier, Signal Generators and DAQ are also kept in isolated cabinets covered with aluminum foil and ant-EMI fiber canopies. All possible measures are taken to minimize EMI and RF noise in our test procedures and in the acquisition of preliminary results reported here.

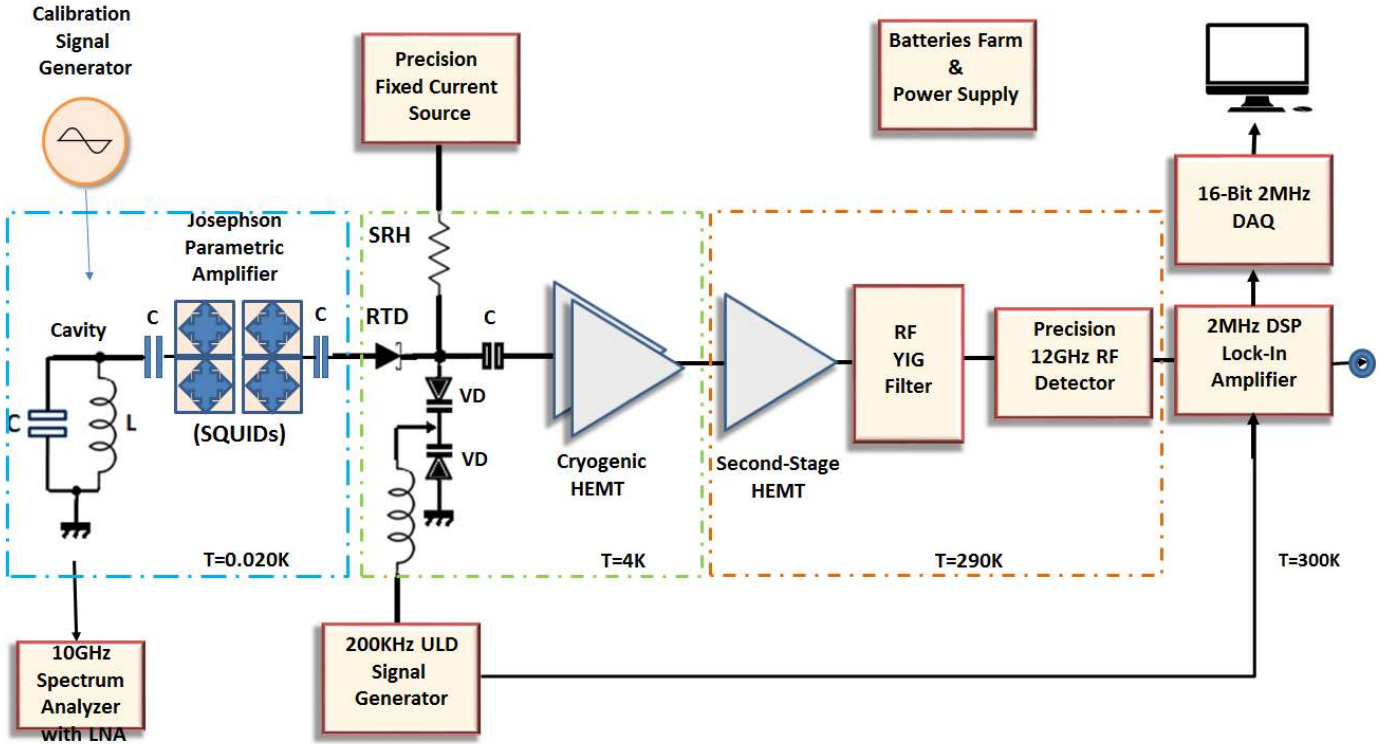

Figure 8: The electronics and instrumentation involved with the proposed axion experiment. The cavity is shown as a lumped LC network which is critically coupled to a sensing antenna (housed externally in a waveguide). The cavity is also weakly coupled (when the measurements are not carried out) to a high-frequency RF source (Agilent 8474E) and a PC-based Spectrum Analyzer (Aaronia GmBH) for calibration and test signal injections. The sensing antenna is madein the form of a tuning fork topology, developed using gold plated 24SWG copper wires. The signals sensed by the antenna are conveyed to a Josephson Parametric Amplifier (JPA) made with two SQUIDs (as explained in detail in the text), which serves as the first amplification stage. The *ac*-coupled output of the JPA is further amplified by a high-gain amplification block, designed around an RTD (Resonant Tunneling Diode) and a pair of matched HEMT MMIC's, a set of LNF-LNC-1.5-6A 1.5-6GHz cryogenic LNAs, with noise temperature of around 2.1K (~0.031dB noise figure) and a gain of 26dB, working at 2K. Since the signal has not reached the sufficient strength to be read out by the RF detector, a room temperature LNA stage follows, comprising a 60K noise temperature HEMT amplifier (Pasternack PE15A1010 2-6GHz LNA 0.9dB NF, 40dB gain LNA @ 290K), which is coupled to an RF detector stage (via a YIG RF filter). This stage consists of a Low-barrier Schottky type power law detector (Agilent/Keysight*8471E* 0.01-12 GHz Planar-Doped Barrier diode detector) which produces a stream of voltages corresponding to the detected RF power from the cavity. The detected RF signal is conveyed to a 2MHz DSP Lock-in Amplifier (Stanford Research *SR865A*) for noise minimized phase-sensitive detection which generates *dc* voltages corresponding to the measured power, which are in turn conveyed to a computer for storage and analysis in a digitized form via a commercial high-precision 16-bit 1MHz Data Acquisition (DAQ) system, which also performs online FFT analysis on the measured samples. A precise and low-distortion 150KHz (to be upgraded in future experiments to 1MHz for higher pulse-width) carrier signal (with modulation index $A_{sig}/A_{mod} \leq 1/2$) is generated by a Stanford *SR260Ultra-Low-Distortion Signal Generator* and given to both the LIA as well as to the cavity sensing and routing circuit. This is a specially designed cryogenic circuit, named *Sensing and Routing Hub (SRH)*,

**analogous to a bias Tee but with added functionality. It serves the purpose of connecting the cryogenic HEMT output to the RTD and in turn to the second stage LNA (which is kept at 290K temperature by cooling with a *Peltier* cooling system) while capacitively coupling the carrier modulation signal to the *ac* line, as well as providing an isolated bias voltage to the RTD coming from a precision voltage source (designed around a precision current source device from Maxim semiconductor). For probing higher axion mass values, we plan to employ Caltech Microwave/Cosmic Microwave CIT112/CIT412 and Low-Noise factory LNF-LNC-23-42WB HEMT's, which are precision cryogenic low-noise amplifiers, with 1 to 12 GHz and 23 to 42 GHz operational regions, respectively, to probe axion masses till around 112.5μeV. For detection purposes at these higher frequencies, an Agilent/Keysight*8474C/8473D* Planar-Doped Barrier Diode Detector (0.01 - 33 GHz) will replace the current 12GHz RF detector.**

The first stage in signal processing and acquisition cascade comprises a special tuning fork antenna coupled to the cavity (shown here as a lumped LC circuit) which is in turn coupled to a narrow-band microwave Josephson Parametric Amplifier (JPA) [14] comprising a network of two (or possibly four) dc SQUID's as per our preliminary design. An external waveguide housing the antenna and connected to the cavity may be incorporated for impedance matching between the two.

A Josephson Parametric Amplifier (JPA) [14] is a non-linear phase-sensitive amplification device which carries outparametric amplification of an ultra-weak signal around its resonance frequency by varying one of its parameters. Since a JPA offers one of the most sensitive non-dissipative detection regimens at the single-photon level available today for ultra-weak microwave detection, it is the most feasible, and at the same time mostcritical stage, of the detection scheme and pivotal to its success. The particular design we propose to employ is a *'degenerate'* Josephson amplifier design [15], an amplifier which could possibly work with the true quantum character of radiation, by the virtue of its ability to work with squeezed states. Such a device is useful in amplifying a signal by adding less quadrature of intrinsic noise to its output than that of the intrinsic electromagnetic fluctuations in the quantum vacuum, while at the same time providing sufficient gain.

The parametric amplifier we contemplate is designed around a pair of Superconducting QUantum Interference Devices (SQUID's), less than one wavelength apart from our center frequency, and each comprising two Josephson junctions, similar to the designs reported earlier [16, 17]. Such designs have claimed an efficient 42% squeezing of 4.2 K thermal noise with a noise temperature of 0.28 K and a gain of around 12 dB. Another viable design in this direction is one reported elsewhere [18]involving a Josephson Traveling Wave parametric amplifier (JTWA) architecture that offers high-gain and low-noise detection at the level of single photons. All these ideas and technologies seem to be a good starting point towards making an ideal weak-measurement, quantum-limited and squeezed state single-photon parametric amplifier for axion detection. Although at this point it is not possible for us to develop a highly-advanced four-SQUID single photon detection amplifier, such a facility would be extremely valuable. The JPA design is in progress and has not been implemented in the experiment so far.

The signal obtained from the JPA (or the tuning fork antenna at the moment) is *ac*-coupled to a carefully-designed amplificationcircuit made around a Resonant Tunneling Diode

(RTD) and some passive components before conveying the signal to the HEMT amplification stage. So far, we have successfully employed commercially available RF tunnel diodes at the LNA input as a low-noise resonance detection and amplification device [13], owing to their negative conductance region, high-speed amplification, low-temperature operation and fast resonant tunneling of electrons, however, in the next stage we propose to use an RTD coupled to the HEMT stage.Tunnel diodes and RTD's [19] have been used in high-frequency microwave amplification, with potential for utility up to mm wave (THz) range. Recent studies involving quantum barrier RTD structures, fabricated using $Al(x)Ga(1-x)As:GaAs:Al(x)Ga(1-x)As$ materials, support frequencies from 5 to about 35GHz [19], which coincides with our region of interest. Novel GaAs/AlGaAs RTD heterostructures with higher sensitivities have been reported in literature which involve a quantum well in the middle [20]. Integrated RTD-HEMT devices have also been proposed by incorporating an HEMT junction with electron injection from a RTD structure at the gate, whereby resonant-tunneling assisted propagation and amplification of plasmons was achieved [21]. We propose a similar custom-made structure to be incorporated in our design to achieve a gain of around 30-40dB using a combo cryogenic RTD-HEMT device.

Working in tandem with TD/RTD, a number of extremely high-Q and low-loss microwave-grade passive components, including a pair of GaAs varicaps (Varactor Diodes), capacitors and inductors are an important part of this scheme, which we have so far achieved success with in detecting weak microwave signals around 1-5GHz range of frequencies.

Following the RTD stage, a pair of matched cryogenic ultra-low-noise amplifiers, working in parallel, further amplify the acquired signal while adding negligible noise. The second stage consists of a traditional room-temperature High-Electron Mobility Transistor (HEMT) (which are a kind of Heterogeneous Field Effect Transistors, or HFET's)Low-Noise Amplifier (LNA) device connected to a commercial YIG *(Yttrium-Iron-Garnet)* RF band-pass filter which provides a band-pass filtered output to an RF detector, which together endeavor to provide highest possible gain (while keeping noise temperature to minimum) to considerably increase the signal power to be detected by the detection scheme.Both the cryogenic as well as the room-temperature low-noise-amplifiers are based upon the usual HEMT architectures, implemented in the form of commercial Monolithic Microwave Integrated Circuits (MMIC's), the working details and designs of these devices in microwave amplification and resonance detection have been discussed elsewhere [13]. Since the quantum limit happens at around 48mK/GHz, our amplification scheme presents about 259mK quantum noise temperature for probing an axion mass window of around 22-23μeV.

The next stage after amplification and filtering is detection of microwave signals, based upon a commercial RF detector. This detector is a high-precision square power law detector of Schottky Barrier Diode (SBD)-monolithic architecture, offering high responsivity and low noise. RF detectors made with barrier diodes offer high sensitivity Noise Equivalent Power (N.E.P.) and they are a set of devices which can detect the lowest detectable microwave signal power. We employ here an extremely sensitive and high-precision GaAs Planar Doped Barrier Diode detector by Agilent (8474E).

Following the amplification and detection stages, the third stage comprises a phase-sensitive detection and low-pass filtering of the signal designed around a Digital Signal Processing (DSP) Lock-In Amplifier (Stanford Research SR865).The basic working principle of a phase-sensitive detection [13] based upon a Lock-in Amplifier, is to mix (multiply) a time-varying noisy signal with a modulation signal at a carrier frequency, and carrying out a Fourier analysis of the modulated (multiplied) signal and selecting only the in-phase components which appear at the carrier frequency. The phase-sensitive detector decomposes a time-varying voltage signal [V(t)] into an X (*In-phase*) and Y (*Quadrature*) component around a carrier frequency ($\omega_c$) as;

$$V(t) = V_0\{X(t)\cos(\omega_c t) + Y(t)\sin(\omega_c t)\} \qquad (21)$$

And produces at its output a voltage corresponding to the product of two sinusoidal waves, one the measured signal (at the center frequency $\omega_0$), and the other the reference signal (at the carrier frequency, $\theta_{ref}$);

$$V_{PSD}(t) = V_0\{\sin(\omega_0 t + \theta_0)\sin(\omega_c t + \theta_{ref})\} \qquad (22)$$

Which after filtration through a Low-Pass filter leaves a clean dc voltage of maximum amplitude proportional to the signal at every point where the signal frequency ($\omega_0$) matches the carrier frequency ($\omega_{ref}$), as;

$$V_{PSD}(t) = \frac{1}{2}V_0\{\cos(\theta_0 - \theta_{ref})\} \qquad (23)$$

In a similar fashion, we multiply the measured signal from the cavity amplifiers with an ultra-low distortion 150KHz carrier frequency signal and then subsequently time-average each in-phase sample produced from the LIA output along the time axis and obtain a time averaged series of *dc* voltages which reproduce the measured signal. Thus, a clean signal average is obtained with substantial SNR improvement and noise reduction. Effectively, one can expect an SNR improvement of around 5 times or more for a raw signal which had poor SNR before. This is the benefit of phase-sensitive detection and time averaging, a quite rudimentary technique in signal analysis.

Monitoring of cryogenic as well as second-stage LNA temperatures and various voltages are carried out using commercial thermometers and digital multi-meters, respectively.

A parallel chain of a commercial low-noise high gain amplifier (RFBay, inc. LNA3035 LNA module, 40dB gain, 0.5-0.9dB NF) connected to a 6GHz spectrum analyzer(Aaronia GmBH HF6060v4), is used for calibration and monitoring purposes with its source a signal from an Agilent (Keysight E8251A) 20GHz RF Signal Generator (via a 20dB attenuator). The injected test signals are obtained from either the same signal generator or from external RF sources of low intensity weakly coupled to the cavity and at times outside the cavity.

It is important to note here that once the JPA is incorporated in the design, it has to be kept in an isolated magnetic-free region for its proper functioning and thus a mu-metal box is made for this unlike the enclosures for other components (which are regular stainless steel cases with SMA connectors).

*3.2 Noise Registry:*

The total noise temperature of the system is calculated as per a simple expression:

$$T_{sys} = T_p + T_c + T_1 + \sum_{i=2}^{n}\left[\frac{T_i}{\prod_{j=2}^{i-1} G_j}\right] \qquad (24)$$

Here, $T_p$ is the system's physical temperature and $T_c$ is the cavity temperature which takes into account a black-body radiation form of the enclosed quantum vacuum (including the quantum fluctuations at $\frac{1}{2}\hbar\omega$). $T_1$ is the noise temperature of the first or primary amplifier (which is the JPA in our case), $T_i$ are the noise temperatures of secondary amplifiers (i.e. the cryogenic and room-temperature LNA's) and $G_j$ are their gains, respectively. For the JPA, once fully implemented, a total noise temperature of around 1.0-1.2K is expected (operating at 20mK) while obtaining a gain of around 15-25dB. With this value, we estimate the total noise temperature of the system not exceeding 5K (excluding the room temperature stages). The individual noise temperatures of each stage, the gains and overall noise overhead of the system are tabulated in Table II.

| Component | $T_{phys}$ (K) | $T_n$K (%) | Gain (dB) |
|---|---|---|---|
| Cavity | 0.020 | 0.020 (0.2) | 0 |
| JPA | 0.020 | 1-1.2 (2-4) | 15-25 |
| cHEMT | 2.0-4.0 | 2-3 (3-5) | 26 |
| RTD | 2.0-4.0 | 0.5-1.0 (1) | 5-10 |
| HEMT | 290K | 60-65 (84-80) | 40 |
| Cables | Various | 7-8 (10) | -10 |
| | | | |
| Total | | ~70-80 (~100) | 76-91 |

**Table II: System noise registry. The left column enumerates the working physical temperature (T$_{phys}$) of each component whereas the middle column depicts the noise temperature (T$_n$) of each component in the system and respective percentage of the overall noise overhead. The total overhead is the system temperature (T$_{sys}$). The far right column gives an estimate of the gain at each stage. Note that the RTD gain listed is the gain obtained while operating in its negative conductance region and in the wake of resonance.**

Another important factor in noise registry, even at the low temperature regime (~mK) is the thermal excitation partition noise, or photon-assisted shot noise, owing to the photonic excitation of electrons emitted from a thermal bath [22]. It has been reported elsewhere [23] that approximately 8-10% of the noise associated with the measurement of a quantum system at sub-

Kelvin physical temperatures is of this kind, having a form $\sim 2(k_B T/h\nu)J_l(1)^2$ with $J_l$ a Bessel function, and may be reduced by a factor of three by using optimal filtering and increasing the thermal conductance of the cables involved.

We plan to continue our tests at liquid nitrogen temperatures of around 70K, but once we acquire a dilution refrigerator and JPA, the antenna and parametric amplifier shall be mounted on the mixing chamber stage at around 20mK(in orderto achieve the quantum limit of $\mathbf{T_{sys}} \sim \mathbf{m_a}$ for 5.4GHz), whereas the SRH and post-amplification stages are at the top chamber of the dilution refrigerator at a temperature of 4K. For the resonant frequency of 5.4GHz, the Johnson noise at 0.020K ambient temperature and a bandwidth (*Δf*) of 20KHz is estimated around $10^{-21}$W (using $Pj = k_B T \Delta f$). In terms of the overall measurement accuracy, we expect a total systemic uncertainty of around 10%.Other than effectively minimizing the thermal noise to a nearly negligible order, the other main concern in terms of the total noise registry of the measurement system is the quantum noise associated with high frequencies involved. This kind of noise is inherent with the detection circuitry, which mainly includes the detector and amplifier, and in low-temperature measurements is known to dominate the total noise overhead. It has been proposed elsewhere to use either single-photon detectorsor phase-sensitive parametric amplifiers working beyond the Standard Quantum Limit to minimize this [24, 25]. However, in addition to great benefits, they have serious shortcomings and limitations as well, such as intrinsic noise arising from zero-point energy fluctuations which are amplified along with the signal excitations, etc. We propose a similar scheme with a new improvement in the read-out stages by incorporating phase-sensitive detection, which we believe could extend a benefit to overcome the noise inherent with the measurement and compensate for the noise backlog.

## 4. Preliminary Tests

Over a period of two months, various tests were carried out based on the scheme proposed in this report, working with both room (~300K) and $LN_2$ (~70K) physical temperatures, to investigate the resonance detection and measurement capabilities for anartificially injected axion signal, especially the noise profile of the signal.

In order to assess the noise profile of our cavity and measurement scheme, Figure 9 illustrates the noise raw Power Spectral Density (PSD) in the region of interest of the noise measured from the cavity without any kind of stimulation and in stringent shielding conditions. The outcome is quite satisfactory showing quite small fluctuations in the spectral distribution of measured power.

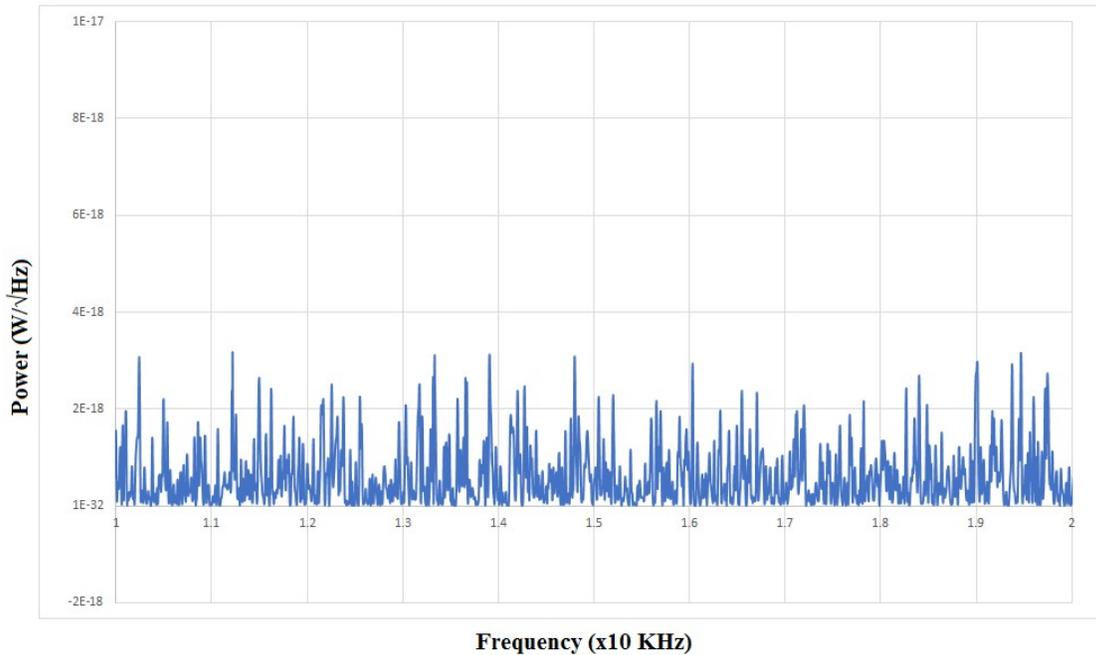

**Figure 9:** Raw (unintegrated) Noise Power Spectral Density (PSD), or the spectral density of voltage fluctuations of the measured noise in the cavity (without any stimulus), for the region of interest in our *dc* measurement scheme (10KHz-20KHz). The result is quite satisfactory with deviations of acceptable magnitude in the measured noise. The sensitivity of our scheme would increase by a few orders once we venture down to the regime of mK temperatures.

Figures 10 and 11 illustrate the time-domain detection of 2.5-2.7GHz Lorentzian-shape signals artificially injected in the cavity as detected by the scheme depicted in Figure 8, excluding the JPA. Signal was picked with the help of our tuning fork antenna mounted in the RF cavity and coupled to the RTD and sensing and routing hub, all kept at $LN_2$ temperatures, and amplified further by 290K HEMT and RF detector stages. Figure 10 is an instantaneous raw signal as received and displayed by the DSO, thus depicting a true representation of a resonance detection event as registered in the cavity, whereas Figure 11 is a processed form of the signal as generated by the LIA and DAQ, thus depicting higher signal to noise ratio.

Figures 12 and 13 are the frequency-domain spectra of the injected signals as detected by the measurement scheme and produced by our 16-bit DAQ, using its Fast Fourier Transform (FFT) spectroscopy software routines. The data received as a function of time is parsed through an online FFT, with a sampling rate of 1MHz (which is sufficient for the analysis of our signal, which bears its maximum component at 300KHz, as per the Nyquist criterion), producing a data set of power as a function of frequencies, which is saved for later analysis and plotting. Based upon the time-domain voltages data in Figure 11, the Figure 12a depicts the 150KHz carrier frequency peak whereas Figure 12b depicts its first harmonic, at approximately 300KHz.

Figure 13 illustrates an overall accumulated view of various batches and data sets depicting detection of resonance within the cavity corresponding to an artificially injected

11.2μeV axion signal, after phase-sensitive detection, filtering and integration, showing a remarkable capability to detect resonance as well as a conspicuous improvement in SNR.

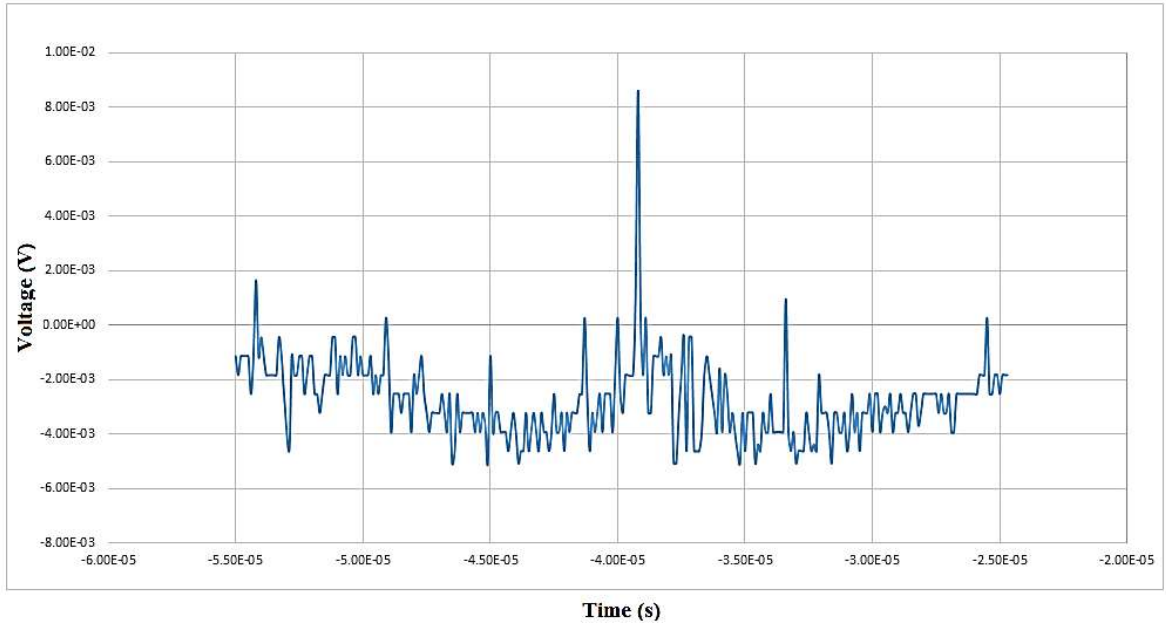

**Figure 10: The form of a measured raw signal as detected by the RF detector in time domain, without any signal processing, to illustrate the form of measured voltage from the cavity. The peak shown in this plot corresponds to the onset of an injected false axionic event (in the form a weak intensity Lorentzian shape signal remotely coupled to the cavity). The calculated raw SNR for this signal was 6.96.**

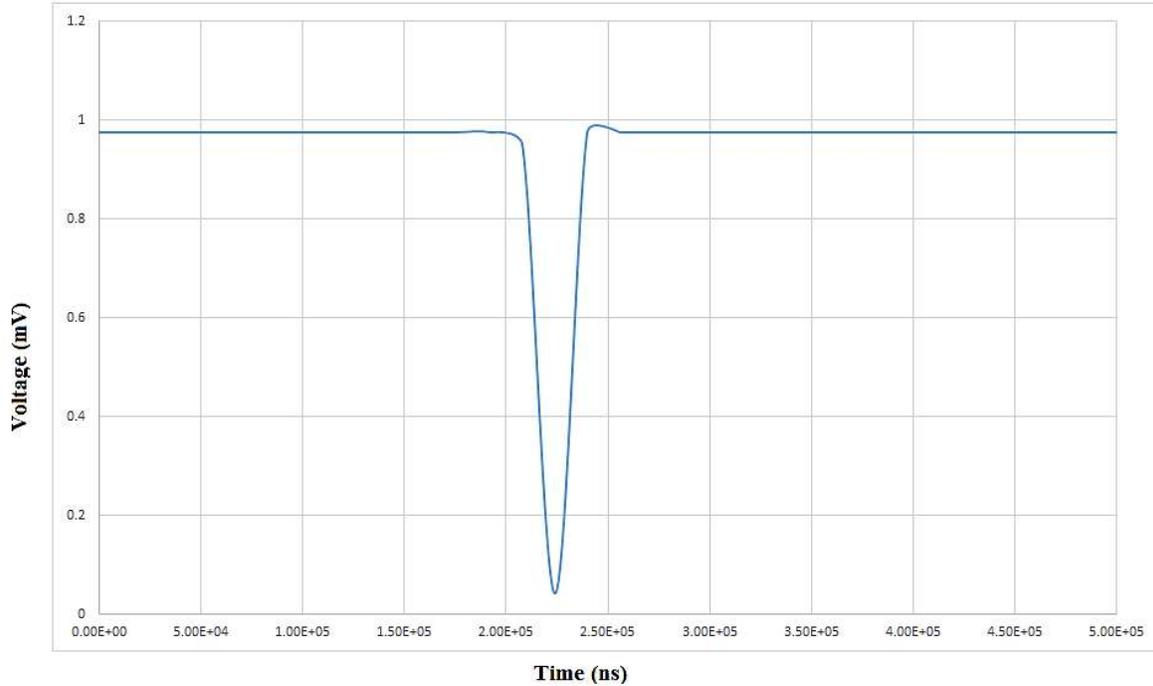

Figure 11: Resonance detection from our scheme. The plot depicts the Phase-Sensitive Detected (PSD) *dc* output from the Lock-In Amplifier (LIA), as obtained by the DAQ and analyzed by the analysis software, as a result of an artificially injected 11μeV signal in the form of a high-frequency 2.5GHz signal from our RF source tapped close to the cavity. The RF resonance detection ability of our PSD scheme can be appreciated from the magnitude and form of this signal.

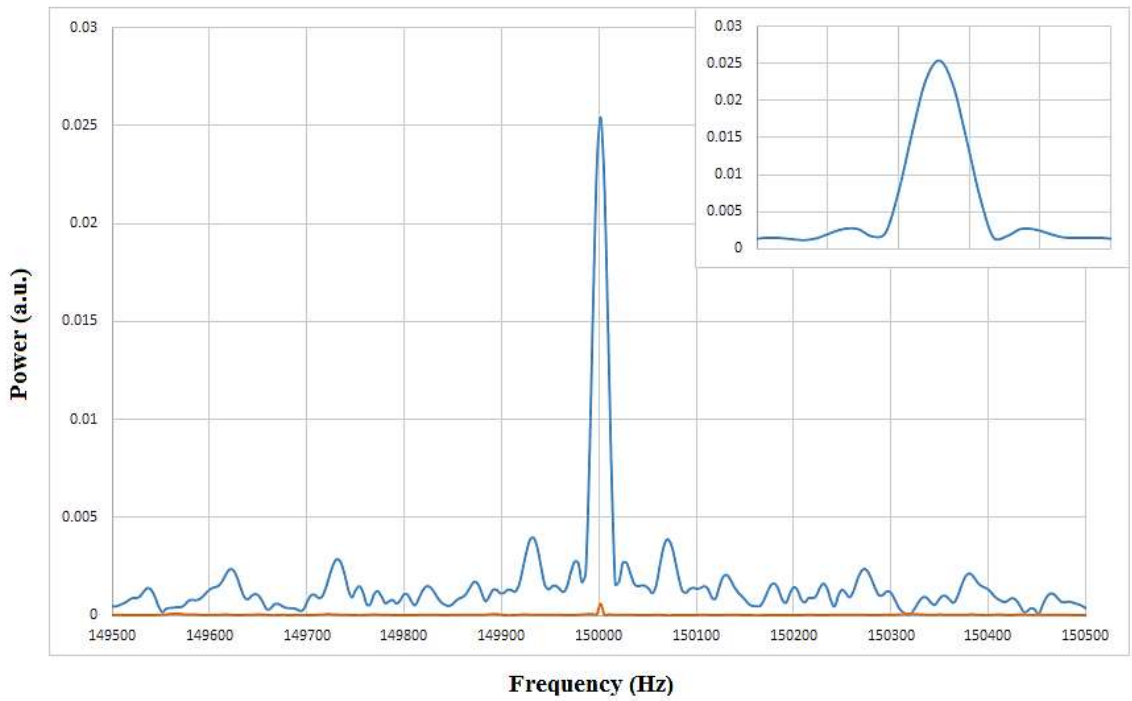

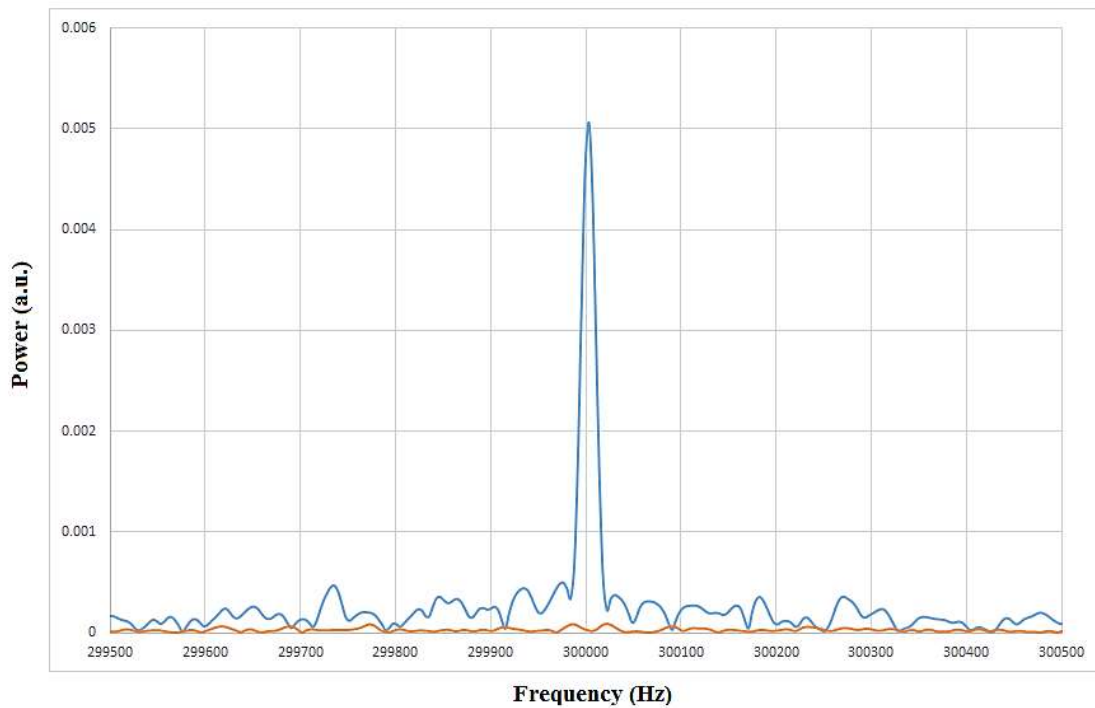

**Figure 12a and b: Illustration of resonance detection in frequency-domain. The detected signal primary and secondary peaks are illustrated in blue color, whereas the noise, in the absence of any injected signal, is illustrated in amber color at the bottom. The inset in Fig 12(a) shows a magnified**

(narrow spectral-range) view of the primary peak, which illustrates the Lorentzian shape of the signal, as expected in resonance. The spectra were constructed from a FFT algorithm using two time-domain data sets from the same batch of experiments, carried out within a short span of seconds, one without any stimulus and one with a weakly coupled artificially injected signal corresponding the axion mass of 11μeV near the cavity (similar to Figure 11).

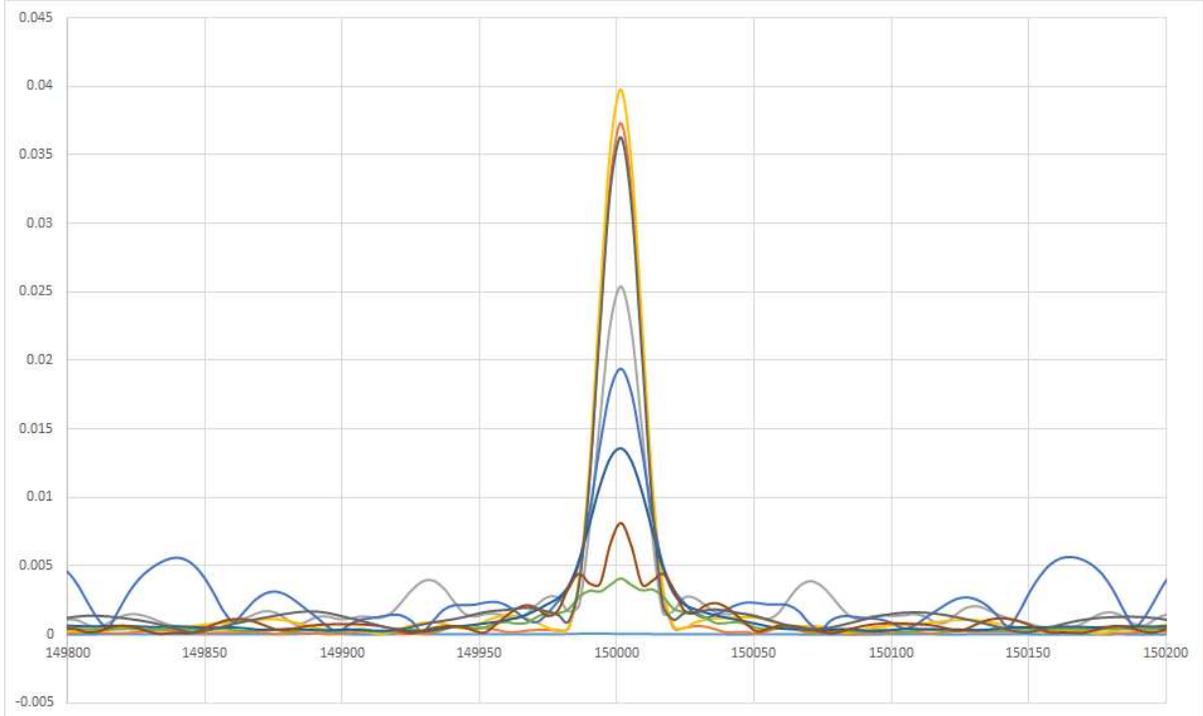

Figure 13: An illustration of accumulated, integrated and processed, spectra from eight different data sets corresponding to detected test resonant signals in cavity, on the lines of Figures 12 b. The calculated SNR from these and similar events ranges from approximately 7 to 9, which illustrates the efficacy of resonance detection in the cavity corresponding to a fixed mass axion value.

## 5. Conclusions and Discussion

With the help of the reported scheme in this report and preliminary tests carried out during a span of two months, we have demonstrated the viability of the proposed scheme especially in the resonance detection for the axion masses of interest in a resonant cavity setting. We have obtained a lowest signal detection capability (at liquid Nitrogen temperatures ~70K) of around $10^{-18}$ to $10^{-17}$W with a raw signal-to-noise ratio of around 3 (and processed SNR of around 10) for frequencies from 2.5 to 5.4GHz, corresponding to axion masses of around 11.2 to 22.5μeV. Once we integrate the signals over long integration periods we can improve our lowest signal limit to around $10^{-22}$W with SNR of around 10 for the full region of our interest (axion mass up to 112.5μeV). Going down to cryogenic temperatures of 20mK and incorporating a JPA, one could confidently venture down to the measurement capabilities of DFSZ signals down to $10^{-25}$W with around 5σ confidence level, using the scheme proposed here.

Caldwell *et al.* (MADMAX collaboration)[26] have reported measuring a $10^{-21}$ W signal at a center frequency of 17 GHz using an ordinary HEMT amplifier operating at room temperature (witha 6σ confidence level) in a one week run, with expectations of 100 time better performance at cryogenic temperatures. In a similar manner, we hope to improve our performance of measuring a $10^{-19}$ W signal manifold by adding a JPA and cryogenic HEMT, as proposed, and going down to sub-Kelvin temperatures.

So far, we have not taken care of the direction of axion wind and the geographical orientation of our measurement system, however, in actual implementation and data taking, the experiment will have to be oriented as per the flux of incident *'axion wind'*. Hence we propose a movable longitudinal detector cavity, which can be oriented in any polar or azimuthal angle, as opposed to a fixed vertical standing cavity. It is important to consider that the "axion wind" or the flux of incoming axions from our galactic halo or other sources, would have a slight spectral shift (*Δf/f*) in its manifested frequency, owing to a velocity dispersion. Axions have been suggested to follow a Maxwell-Boltzmannian velocity distribution [2]. There would also be a modulation and anisotropy in its flux over space and time. Thus, an intelligent search would accommodate these factors in detection and analysis, in addition to taking care of detector location and data taking times.

We would like to mention here two points of extreme importance for an actual axion search, which have been overlooked in the results presented in this paper, however would be implemented in the next stage of our experiments. The first of these is the line width of the detection. It would be extremely important to obtain a line width of approximately 1MHz in the frequencies of proposed mass range, to obtain a peak covering the full span of cavity centered at the cavity center resonance frequency. While insofar we have used a 150KHz line width (owing to the limitations of our ultra-low-distortion signal generator frequency), our next stage of experiments would be based on a full 1MHz cavity line width. The second factor is the shape of *"synthetic axion"* injected signal, as coupled to the cavity to simulate the detection for test and calibration purposes. We have incorporated a simple Lorentzian shape signal injection scheme corresponding to the axion mass of interest, however an actual simulation would entail a synthetic axion signal shape following thecosmological axion distribution, *i.e.* a Maxwellian spectral distribution.

While proposing a detection scheme and an experiment to detect axions of specific fixed mass, we also highlight the possibility of axions with a variable mass distribution, suggesting that axions or axion-like particles may exist in a range of masses, thus having a variable energy content similar to photons. Based upon our calculations and simulations we suggest an axion mass distribution of 56.25 to 146.25 μeV (corresponding to a frequency range of 13.5 to 35.1GHz) centered at a mean value of 101μeV + 1.0μeV. Experiments will be needed to assess this assumption.As per various theoretical axion models, whether scalar, pseduoscalar or vector dark matter models, axions over a large mass range are possible [27]. Variable mass scheme for cold dark matter, like axions, have been suggested as a plausible mechanism which even favors inflation coupling [28].

There have been suggestions of values close to our proposed value by various theoretical studies during the last one decade. As per Shellard and Battaye [3], with a hubble parameter value around 0.5, the string theory-based estimates suggest a value of axion mass around 100µeV. Ballesteros in their review paper [29] have suggested an axion mass of 100µeV. In two separate studies [30, 31] an axion mass range has been suggested around 50-1500 and 50-200µeV, respectively. A narrower estimate was suggested by Kawasaki et al [32] around 60-150µeV. It has been suggested elsewhere [33] that the recent high energy scale of inflation, as indicated by the latest BICEP2 results, puts a lower bound on the axion mass around 70µeV. All these estimates support our proposed axion mass range from 22.5µeV to 111.5µeV, which is within reach of current technology available with small research laboratories like ours.

On the other hand, based upon the analysis of five different condensed matter experiments data involving a special kind of Josephson junctions, arguments by Beck [34] suggest a mass range of 104 to 110µeV, not far from our suggested value. In addition, he also suggests the need for more experiments to assess these propositions, especially taking into account the annual modulation (of around 10%) in the Earth's velocity with respect to the galactic halo, which has a peak in June and minimum in December.

Considerations of important factors encountered by a terrestrial axion detector, such as daily modulation of incident (anisotropic) axion flux and the signal amplitudes from our galactic halo (or other celestial sources), the direction of incidence as well as the rotation of Earth and our solar system with respect to the sources of axion wind, as discussed in detail elsewhere [35, 36], are quintessential elements of a realistic axion detection experiment design, just as spectral broadening and modulation are, as highlighted in this report and elsewhere [24]. These factors are not only important variables of the experiment but also a touchstone for the validity of a true axionic event of celestial source.

In the end, we suggest that one would have to look beyond the current measurement strategies in order to probe the DFSZ region axions. One such way could be incorporation of quantum well structure-based detectors, such as quantum well-based HEMT's and approaches like *'Qubits'* and other recent advancements in cirQED [37]. However, approaches like Qubits and associated hardware would add significantly to the noise overhead and complications in the experiment hardware, thus it would be imperative to find ways to avoid in increasing the noise overhead. There will have to be consideration of quantum back action [38], especially a measurement approach which takes into the account back action caused by the detector itself on the measurand, and what kind of role it plays in a realistic axion signal detection. Another problem, as pointed out by the Carugno group [39], is the amplification of vacuum fluctuations. Any sensitive amplification scheme to detect axions would also amplify the fluctuations of the quantum vacuum which would pose a background similar to axions. In such events, it would be necessary to incorporate some cosmological cuts and analysis of the detected noise to eliminate the background signal corresponding to vacuum fluctuations, if at all it is possible. Otherwise, an altogether different approach would have to be sought than the customary microwave signal amplification and detection approaches. On the other hand, we argue, what if the fluctuations in the electric field of quantum vacuum can themselves interact with or drive the axion-induced electric fields to higher amplitudes? The quantum fluctuations remain an important factor to consider in axion detection schemes and must not be overlooked.

Similarly, ordinary cylindrical cavity geometries might not be sufficient, novel new methods like re-entrant cavities and torus geometries, which could facilitate resonance effects, could be extremely valuable. These would pose additional problems in setting up the detectors and magnets, which would requireadditional efforts.

An important new direction in axion search experiments would be to find novel new ways for ultra-weak single photon detection, such as photon counting. For instance, as reported in a recent study [40], various cavity frequencies can be mapped onto individual states of a qubit by means of a Microwave Josephson Photomultiplier (JPM), and thus very weak microwave signals can be measured in a cavity, extending the concept of a Photomultiplier tube to single photon detection and counting. Raw single-shot photon measurement fidelity of more than 90% was demonstrated in their experiments while preserving the quantum states of the system in a Quantum Non-Demolition (QND) manner. Such methods could offer a viable solution in detecting ultra-weak photons generated as a result of axions in a microwave cavity setting. Additionally, a qubit placed in a microwave cavity while incorporating the effect of quantum back action and a detection scheme which could detect this back action may be useful. Once again, the added noise and complexities in the experiment would be a significant factor to consider in such schemes.

We suggest here some humble ideas to advance the current knowledge in axion and cold dark matter particle searches. We are confident that once implemented, an experiment devised on these lines as proposed in our scheme, as well as considering the ideas discussed herein, could assist in the detection ofaxions/ALP's with sufficient statistical certainty after a run for few months. Independent of the existence of the dark matter as a plausible explanation of the galactic rotational velocity curve discrepancies and other astrophysical observations, axions have their profound existence as per the QCD and instanton effects. Axions or similar particles do exist even if the dark matter does not, however they may be difficult to distinguish from other excitations in the absence of any cosmological models to map them onto. It is a search worth continuing for our correct understanding of the quantum fields and matter constituting the universe,independent of*a priori* assumption of dark matter.

**Acknowledgments:**

Author acknowledges, with great gratitude, valuable discussions with Prof. Z. H. Shah, Prof. GiovanniCarugno, Giuseppe Ruoso and Nicholas Crescini, which were fruitful in refining the manuscript and ideas presented herein. I am grateful to the support and hospitality by the University of Padua and Carugno group at the INFN Legnaro, Italy, during my one-week long visit there. Funding support from the Deanship of Scientific Research, Jazan University, in setting up my laboratory is acknowledged with gratitude.